\documentclass[12pt]{iopart}

\usepackage{iopams} 
\usepackage{graphics}
\usepackage{graphicx}
\usepackage{hyperref}
\usepackage{float}
\begin{document}

\title[Generalized Finch and Skea model compatible with observational data]{Generalized Finch and Skea model compatible with observational data}

\author{Rinkal Patel, B.S.Ratanpal, D.M.Pandya}

\address{Department of Applied Science and Humanities, Parul University, Limda, Vadodara, 391 760, Gujarat, India, \\ Department of Applied Mathematics, The Maharaja Sayajirao University of Baroda, Faculty of Technology and Engineering,
	Vadodara, 390 001, Gujarat, India, \\ Department of Mathematics, School of Technology, Pandit Deendayal Energy University, Raisan, Gandhinagar, 382426,
	Gujarat, India.}
\ead{rinkalpatel22@gmail.com, bharatratanpal@gmail.com, dishantpandya777@gmail.com}
\vspace{10pt}
\begin{indented}
\item[]August 2023
\end{indented}

\begin{abstract}
	We present an entirely new class of solutions to Einstein's field equations that correspond to a static spherically symmetric anisotropic system by generalizing the Finch and Skea ansatz using the linear equation of state for the gravitational potential $g_{rr} $. Based on physical requirements, regularity condition, and stability, we make various assumptions about the model's parameters. The exact solution generates a plausible model of a compact star PSR J0348+0432 that satisfies all physical criteria. The requirements for a well-behaved compact star are met, including regularity, equilibrium, causality, stability, energy, and compactness limitations.

\end{abstract}

\vspace{2pc}

\section{Introduction}

	The study of compact stars has piqued the interest of researchers working on relativistic astrophysics. Various approaches have been opted to model compact stars including isotropic fluid, anisotropic fluid, charged isotropic and charged anisotropic fluid inside the compact objects. However, obtaining exact solutions for isotropic scenarios is more difficult than for anisotropic solutions. Furthermore, in extremely dense conditions, pressures may bifurcate into radial and tangential components, resulting in pressure anisotropy. When solving problems with compact stars in General Relativity(GR), it is common practice to assume objects with spherically symmetric and isotropic properties. However, the isotropy and homogeneity of astrophysical compact stellar objects may ideally have solvable features, but they do not have to be general physical stellar object characteristics. As a result, fluid pressure may have two distinct components that are responsible for providing an anisotropic factor ($p_{r}-p_{\perp}$) where inhomogeneity results from radial pressure ($ p_{r}  $) and tangential pressure ($ p_{\perp} $) and thus may deprive the internal system of matter distribution of an idealized version of isotropic case. Ruderman \cite{ruderman1972pulsars} first mentioned this idea in his extensive review work on pulsar structure and dynamics in the nuclear regime when matter distributions have a high density. Bowers \cite{bowers1974anisotropic} have extensively discussed the various causes of anisotropy's appearance. Since then, a number of research articles incorporating anisotropy in pressure have appeared in the literature. Anisotropy arises in self-gravitational compact stars due to the occurrence of exotic phase transitions by 
	Sokolov \cite{sokolov1980phase}, electromagnetic fields, rotations, superfluid or type-A fluids by Kippenhahn et al.\cite{kippenhahn1990stellar}, pion and meson condensations by Sawyer \cite{sawyer1972condensed}, core formation, and other phenomena by Herrera and Santos (\cite{herrera1997local}, \cite{herrera39santos}),  Herrera and Leon \cite{herrera1985isotropic},  Dev and Gleiser (\cite{dev2002anisotropic}, \cite{dev2003anisotropic}), Cosenza et al.\cite{cosenza1982evolution}, Bayin \cite{bayin1982anisotropic}, Krori et al.\cite{krori1984some}, Ponce \cite{ponce1987new}, Bondi \cite{bondi1992anisotropic}, Herrera et al.(\cite{herrera2001conformally}, \cite{herrera2009expansion}, \cite{herrera2008all}, \cite{herrera2010cavity}, \cite{herrera2012dynamical}), Mak and Harko \cite{mak2002exact} have extensively investigated highly dence objects. This implies that self-gravitational systems have two types of pressure components, the radial component ($ p_{r}  $) and the tangential component ($ p_{\perp}  $). As a result, radial and tangential pressures become unequal, giving rise to the concept of local anisotropy in the study of self-gravitational fluids. Dev and Gleiser \cite{dev2002anisotropic} have also presented new exact solutions for the distribution of matter with tangential pressures and uniform energy density, as well as a proposed equation that relates the radial and tangential pressures in order to integrate analytically the Einstein field equations. Herrera and colleagues (Herrera and Barreto \cite{herrera2013general}, Herrera et al.\cite{herrera2014conformally}) investigated static anisotropic stars in this context by investigating the effects of Newtonian and general relativistic regimes. The effect of anisotropy on the physical properties of a distribution of matter in relativistic compact objects has been studied by authors such as Herrera and Santos \cite{herrera39santos}, Herrera and Leon \cite{herrera1985isotropic}, Dev and Gleiser \cite{dev2003anisotropic}, Cosenza et al.\cite{cosenza1982evolution},Bayin \cite{bayin1982anisotropic}, Krori et al.\cite{krori1984some}, Ponce \cite{ponce1987new}, Bondi \cite{bondi1992anisotropic},  Herrera et al.(\cite{herrera2001conformally},  \cite{herrera2008all}),  Mak and Harko (\cite{mak2002exact},  \cite{mak2003anisotropic}), Viaggiu \cite{viaggiu2009modeling}.

An equation of state(EoS) describes the state of matter under a specific set of physical conditions such as pressure, volume, temperature or internal energy. EoS can be used to describe the properties of pure substances and mixtures in liquids, gases and solid states as well as the state of matter inside stars. EoS is also used to model the density of matter in the interiors of stars including neutron stars. Numerous previous works in the literature used general barotropic equations of state in which density and pressure are related in linear, quadratic or polytropic form. Without a doubt, selecting a suitable EoS for the demonstration of astrophysical compact objects is a fundamental issue. Mak and Harko \cite{mak2002exact} deliberated the utility of linear equations of state in modeling static spherically symmetric anisotropic quark matter distributions. Sharma and Maharaj \cite{sharma2007class} used linear equation of state to build relativistic models that are compatible with observational data. Thirukkanesh and Maharaj \cite{thirukkanesh2008charged} investigated three classes of new exact solutions to the Einstein-Maxwell system by having different gravitational potential Z and electric field intensity E. Takisa et al.\cite{mafa2014charged} studied an exact solution to the Einstein-Maxwell system for a charged anisotropic compact body in the linear regime.  Since then, a number of articles incorporating linear equation of state in terms of  pressure and density have appeared in the literature like Harko and Mak \cite{harko2016exact}, Hansraj and  Banerjee \cite{hansraj2018equilibrium}, Ivanov \cite{ivanov2020linear}, Rej and Bhar \cite{rej2021charged}, Das et al.\cite{das2022anisotropic}, Prasad and Jitendra \cite{prasad2022anisotropic},  Patel et al.\cite{patel2023new}. For physically viable relativistic models of compact stars, Ngubelanga et al.\cite{ngubelanga2015compact} used linear equation of state in isotropic coordinates. Varela et al.\cite{varela2010charged} thoroughly investigated the importance of an equation of state in a stellar model using Krori and Barua's method in terms of pressure anisotropy and linear or nonlinear equations of state. Because of the increased nonlinearity in the field equations, models with a quadratic equation of state are relatively rare Feroze and Siddiqui \cite{feroze2011charged}, Maharaj and Takisa\cite{maharaj2012regular}, Sharma and Ratanpal \cite{sharma2013relativistic}, Malaver \cite{malaver2014strange}, Malaver and Kasmaei \cite{malaver2020relativistic}. The polytropic EoS plays an important role in the composition of astronomical compact star models in explaining the various features of compact stars. Herrera and Barreto \cite{herrera2013newtonian} conducted a detailed study for Newtonian polytropic models when the fluid is anisotropic. Thirukkanesh and Ragel (\cite{thirukkanesh2012exact}, \cite{thirukkanesh2013class}) illustrated the realistic features of uncharged compact star models in the polytropic EoS regime. Takisa and Maharaj \cite{takisa2013some} have also reviewed the physical implications of polytropic EoS on charged anisotropic compact star models. Recently Singh et al.\cite{singh2022anisotropic} investigated an isotropic solution for polytropic stars in 4D Einstein-Gauss-Bonnet gravity.	

The work organized as follows. In Sec.2, the Einstein field equation describing on spherically symmetric static anisotropic matter distribution is given. We assume particular choice of metric potential $e^{\lambda}$ and linear EoS in Sec.3. In Sec.4, we have solved the system to generate a new model along with the related matching conditions. Whereas the physical viability	and stability analysis of our model with graphical representation for compact star PSR J0348+0432 have been studied in Sec.5 and Sec.6. Finally, concluding remarks have been given in Sec. 7.

\section{The Spacetime Metric}
\label{sec:2}
We write the interior space-time of a static spherically symmetric distribution of anisotropic matter in the form

\begin{equation}\label{e2}
	ds^2 = e^\nu dt^2-e^{\lambda}dr^2 -r^2(d\theta^2 +sin^2 \theta d\phi^2)
\end{equation}
where,
\begin{equation}\label{elam}
	e^{\lambda}=\left(1+\frac{r^2}{R^2}\right)^{n}
\end{equation}

The constant $ \frac{1}{R^2} $  can be identified with the curvature parameter  having dimension of a length and $n > 0$ is a dimensionless parameter. Note that the
ansatz(\ref{elam}) is a generalization of the Finch and Skea (\cite{finch1989realistic}) model which can be regained by setting n = 1. We take the energy-momentum tensor of the anisotropic	matter filling the interior of the star in the form
\begin{equation}\label{e4}
	T_{ij}=(\rho+p_{\perp})u_iu_j+p_{\perp}g_{ij}+(p_r-p_{\perp})\chi_i\chi_j
\end{equation}
where $ \rho $ is the matter density, $ p_{r} $ is the radial pressure, $ p_{\perp} $ is the tangential pressure, $u^{i}$ is the four-velocity of the fluid and $ \chi^{i}$ is a unit spacelike four-vector along the radial direction so that $u^{i}u_{i}= -1, \chi^{i}\chi_{j} = 1$ and $u^{i}\chi_{j}= 0$. With spacetime metric (\ref{e2}) and energy-momentum technique (\ref{e4}), the Einstein's field equations takes the form		
\begin{equation}\label{e6}
	8\pi\rho  = \frac{1-e^{-\lambda}}{r^2}+\frac{e^{-\lambda}\lambda'}{r},
\end{equation} 
\begin{equation}\label{e7}
	8\pi p_{r}  = \frac{e^{-\lambda}\nu'}{r}+\frac{e^{-\lambda}-1}{r^{2}},
\end{equation}
\begin{equation}\label{e8}
	8\pi p_{\perp} =e^{-\lambda} \left(\frac{\nu^{''}}{2} +\frac{\nu'^2}{4}-\frac{\nu' \lambda'}{4}+\frac{\nu'-\lambda'}{2r}\right),
\end{equation}
\begin{equation}\label{e21}
	8\pi \Delta = 8\pi P_{r}-8\pi P_{\perp}.
\end{equation} 

where primes denote differentiation with respect to r. The system of equation (\ref{e6}-\ref{e21}) governs the behaviour of the gravitational field for an anisotropic fluid distribution. Equation (\ref{elam}) and (\ref{e6}) leads to,

\begin{equation}\label{rho}
	\rho=\frac{1-\left(1+\frac{r^2}{R^2}\right)^{-n}}{r^2} + \frac{2n\left(1+\frac{r^2}{R^2}\right)^{-1-n}}{R^2}
\end{equation}
The expression for density $\rho(r)$ is finite at the centre of the star.

	\section{Generating Model}
\label{sec:3}
\noindent
EoS is useful in describing the properties of pure substances and mixtures in liquids, gases and solid states as well as the state of matter in the interior of stars. Many researchers have presented their idea on the linear equation of state. We consider a linear equation of state between the radial pressure $ p_{r} $  and matter density $ \rho $ as
\begin{equation}\label{e14}
	p_{r}=A\rho-B,
\end{equation}
where A and B are constants. The radius of the star with this pressure distribution is obtained by using the condition 
\begin{equation*}
	p_{r}(r=a) = 0,
\end{equation*}

gives,
\begin{equation}\label{e15}
	B = A\rho_{a}.
\end{equation}	 
We substitute equation (\ref{e15}) in (\ref{e14}) , We get
\begin{equation}\label{e16}
	P_{r}=A\rho-A\rho_{a}=A(\rho-\rho_{a}).
\end{equation}
Applying the equation (\ref{e16}) in equation (\ref{e7}), We get 
\\
\begin{equation*}
	\nu'=re^{\lambda}[A(\rho-\rho_{a})-(\frac{e^{-\lambda}-1}{r^{2}})],
\end{equation*}
\begin{equation*}
	\nu'=\frac{(1+A)\left(-1+p_1^{-1}\right)}{r} + \frac{Ar\left(-1+\left(1+\frac{a^2}{R^2}\right)^{-n}\right)}{a^2\left(1+\frac{r^2}{R^2}\right)^{n}}
\end{equation*}
\begin{equation}
	+ \frac{2Anr\left(-\left(1+\frac{a^2}{R^2}\right)^{-1-n}+\left(1+\frac{r^2}{R^2}\right)^{-1-n}\right)p_1}{R^2}, 
\end{equation}
\begin{equation*}
	\nu=Log\frac{C(r^2+R^2)^{An}}{r^{A+1}}-\frac{p_1^{-1}\left(r^2+R^2\right)\left(a^2\left(-1+2n+p_2\right)+R^2\left(-1+p_{2}\right)\right)}{2a^2(1+n)\left(1+\frac{a^2}{R^2}\right)^{n}(a^2+R^2)} 
\end{equation*}

\begin{equation}
	+\frac{(1+A) Hypergeometric2F1[-n,-n,1-n,-\frac{R^2}{r^2}]}{2n\left(1+\frac{r^2}{R^2}\right)^{n}\left(1+\frac{R^2}{r^2}\right)^{n}},
\end{equation}

\begin{equation*}
	e^{\nu}=\frac{C(r^2+R^2)^{An}}{r^{A+1}}exp\left[-\frac{p_1^{-1}\left(r^2+R^2\right)\left(a^2\left(-1+2n+p_2\right)+R^2\left(-1+p_2\right)\right)}{2a^2(1+n)\left(1+\frac{a^2}{R^2}\right)^{n}(a^2+R^2)} \right] 
\end{equation*}

\begin{equation}\centering\label{enu}
	\times \frac{(1+A) Hypergeometric2F1[-n,-n,1-n,-\frac{R^2}{r^2}]}{2n\left(1+\frac{r^2}{R^2}\right)^{n}\left(1+\frac{R^2}{r^2}\right)^{n}}.
\end{equation}

where C is a constant of integration.
\section{Matching Condition}
\label{sec:4}
\noindent  The interior spacetime metric should continuously match with schwarzschild exterior spacetime
\begin{equation}
	ds^{2}=\left(1-\frac{2M}{r}\right)dt^{2}-\left(1-\frac{2M}{r}\right)^{-1}dr^{2} -r^2(d\theta^2+sin^2\theta d\phi^2),
\end{equation} 
at the boundry $ r=a $ of star.
This leads to 

\begin{equation*}
C = \frac{a^{A+1}}{(a^2+R^2)^{An}}	exp\left[\frac{\left(a^2\left(-1+2n+\left(1+\frac{a^2}{R^2}\right)^{n}\right)+R^2\left(-1+\left(1+\frac{a^2}{R^2}\right)^{n}\right)\right)}{2a^2(1+n)}\right]
\end{equation*}

\begin{equation}
	exp\left[-\frac{(1+A) Hypergeometric2F1[-n,-n,1-n,-\frac{R^2}{a^2}]}{2n\left(1+\frac{a^2}{R^2}\right)^{n}\left(1+\frac{R^2}{a^2}\right)^{n}}\right],
\end{equation}

\begin{equation}
	R =\sqrt{\frac{a^2\left(1-\frac{2m}{a}\right)^{(1/n)}}{1-\left(1-\frac{2m}{a}\right)^{(1/n)}}}.
\end{equation}
The expression of matter density, radial pressure and tangential pressure 
takes the form
\begin{equation}
	\rho=\frac{1-\left(1+\frac{r^2}{R^2}\right)^{-n}}{r^2} + \frac{2n\left(1+\frac{r^2}{R^2}\right)^{-1-n}}{R^2},
\end{equation}
\begin{equation*}
	p_{r}=A\left(-\frac{1-\left(1+\frac{a^2}{R^2}\right)^{-n}}{a^2} +\frac{1-\left(1+\frac{r^2}{R^2}\right)^{-n}}{r^2} \right)
\end{equation*}

\begin{equation}\label{pr}
	- A\left( \frac{2n\left(1+\frac{a^2}{R^2}\right)^{-1-n}}{R^2}- \frac{2n\left(1+\frac{r^2}{R^2}\right)^{-1-n}}{R^2}\right),
\end{equation}

\begin{equation}
		p_{\perp}= \frac{2a^2nr^2(a^2+R^2)}{4a^4r^2(a^2+R^2)^2(r^2+R^2)^2p_1p_2}\left(f_1(r)+\frac{\left(f_2(r)\right)^{2}}{p_2^{2}}-f_3(r)\right),
\end{equation}
where,
\begin{equation*}
	f_1(r)=Ar^2p_4 p_1 R^2(r^2+R^2)-a^2\left(p_3p_2R^2(r^2+R^2)\right)
\end{equation*}
\begin{equation*}
	-a^4p_2\left(r^2(-1+Ap_5+p_1)+(1+A)p_3R^2\right)
\end{equation*}	
\begin{equation*}
	+A\left(-r^4p_6p_1+(-1+2n)r^2R^2(p_2-p_1)+p_3p_1R^4\right),
\end{equation*}
\begin{equation*}
	f_2(r)=-Ar^2p_4p_1 R^2(r^2+R^2)	+a^2\left(p_3p_1R^2(r^2+R^2)\right)
\end{equation*}
\begin{equation*}
	+a^4p_2\left(r^2(-1+Ap_5+p_1)+(1+A)p_3R^2\right)
\end{equation*}
\begin{equation*}
	+A\left(-r^4p_6p_1+(-1+2n)r^2R^2(p_2-p_1)+p_3p_1R^4\right),
\end{equation*}	
\begin{equation*}
	f_3(r)=\frac{2a^2(a^2+R^2)}{p_2}\left(p_7-a^4p_2p_8-a^2p_{9}-2a^2(a^2+R^2)(r^2+R^2)p_{10}\right),
\end{equation*}

\begin{equation*}
	p_1=\left(1+\frac{r^2}{R^2}\right)^{n},
\end{equation*}
\begin{equation*}
	p_2=\left(1+\frac{a^2}{R^2}\right)^{n},
\end{equation*}	
\begin{equation*}
	p_3=\left(-1+\left(1+\frac{r^2}{R^2}\right)^{n} \right),
\end{equation*}	
\begin{equation*}
	p_4=\left(-1+\left(1+\frac{a^2}{R^2}\right)^{n} \right),
\end{equation*}	
\begin{equation*}
	p_5=\left(-1+2n+\left(1+\frac{r^2}{R^2}\right)^{n} \right),
\end{equation*}	
\begin{equation*}
	p_6=\left(-1+2n+\left(1+\frac{a^2}{R^2}\right)^{n} \right),
\end{equation*}	
\begin{equation*}
	p_7= Ar^2p_4p_1R^2(r^2+R^2)((1+2n)r^2+R^2),
\end{equation*}	
\begin{equation*}
	p_8=\left(r^4(1+A(-1+2n)p_3+(2n-1)p_1)\right)
\end{equation*}
\begin{equation*}
	+2r^2(1+A(1+n+(n-1)p_1)+(n-1)p_1)R^2-(1+A)p_3R^4,
\end{equation*}
\begin{equation*}
	p_{9}=p_2R^2(r^2+R^2)(r^2(1+(2n-1)p_1)-p_3R^2)+A(-(2n+1)r^6p_6p_1
\end{equation*}
\begin{equation*}
	-r^4R^2(2n(p_1+p_2)-p_2-2p_1+4n^2p_1+3p_2p_1)
\end{equation*}
\begin{equation*}
	+r^2R^2(2n(p_2-p_1+p_1p_2)+2p_2+p_1-3p_1p_2)-p_3p_2R^6,
\end{equation*}
\begin{equation*}
	p_{10}=2a^2nr^2(a^2+R^2)(Ar^2p_3p_1R^2(r^2+R^2)-
\end{equation*}
\begin{equation*}
	a^4p_2(r^2R^2(-1+Ap_5+p_1)+(1+A)p_3) - a^2 p_3p_2R^4
\end{equation*}
\begin{equation*}
	-a^2(p_3p_2R^2(r^2+R^2)+A(-r^4p_6p_1+(2n-1)r^2R^2(p_2-p_1)))
	\end{equation*}
		\section{Physical requirements for well-behaved solutions}
	\label{sec:5}
	\noindent
	The authors of Kuchowicz \cite{kuchowicz1972differential}, Buchdahl \cite{buchdahl1979regular}, Knutsen \cite{knutsen1988some} and Murad and Fatema \cite{murad2015some} have mentoined the set of conditions to verify the model is physically justifiable. These condition have been checked for physically plausible solution for the star PSR J0348+0432.

	\subsection{Regularity Condition:}
	\label{sec:6}
	(i)The metric potentials $ e^{\lambda(r)}\ge 0,\; e^{\nu(r)}\ge 0 $,  $ 0 \le r \le a  $. These features are depicted in Fig.(\ref{fig.1}) and Fig.(\ref{fig.2}).
	The above requirements are met in our model for appropriate parameter selection. From the equations (\ref{elam}) and (\ref{enu}), we can observe that $ e^{\lambda(0)}=1 $ and $ e^{\nu(0)}=constant. $ 
	\\ Again,  $ e^{\lambda(0)}=1 $, indicating that it is finite at the stellar configuration's centre $ (r = 0) $.
	Also, it is simple to see that $ (e^{\lambda(0)})^{'}=(e^{\nu(0)})^{'} = 0 $. This implies that the metric is regular in the center and behaves well throughout the stellar interior, which will be depicted graphically.
	\\(ii) $ \rho(r) \ge 0 ,\;\;\;\;p_{r}(r)\ge 0 ,\;\;\;\;p_{\perp}(r)\ge 0 $ \;\;\;\; for  $ 0 \le r \le a.$
	\\ From Eq.(\ref{rho}), we note that density remains positive for the various values of 'n'.
	\begin{equation*}
		\rho_{(r=0)}=\frac{3n}{R^2},
	\end{equation*}
	\begin{equation*}
		p_{r}(r=0)=	A\left(\frac{p_4}{a^2}+\frac{n\left(3-2\left(1+\frac{a^2}{R^2}\right)^{(-1-n)} \right)}{R^2}\right),
	\end{equation*}
	\begin{equation*}
		p_{\perp}(r=0)=	A\left(\frac{p_4}{a^2}+\frac{n\left(3-2\left(1+\frac{a^2}{R^2}\right)^{(-1-n)} \right)}{R^2}\right).
	\end{equation*}
	We can see that the tangential and radial pressures are still positive at the centre r = 0. Fig.(\ref{fig.3}), Fig.(\ref{fig.4}) and Fig.(\ref{fig.5}) have shown the fulfilment of the requirements throughout the compact object  PSR J0348+0432 for the variour values of 'n'.
	
	(iii) $ p_{r}(r=a) = 0 $
	\\ From the equation (\ref{pr}), We can see that the radial pressure disappears at the boundary $ r= a  $. Which is shown in graphical representation with Fig.(\ref{fig.4}).
	\subsection{Causality Condition:}
	\label{sec:8}
	The causality condition demands that $0 \le \frac{dp_{r}}{d\rho} \le 1$ and $0 \le \frac{dp_{\perp}}{d\rho} \le 1 $ at all interior points of the star. Hence from equation  (\ref{e16}), we have
	\begin{center}
		$ 	\frac{dp_{r}}{d\rho}=A. $
	\end{center}
	
	Eq.(\ref{e14}) shows that since $ dp_{r}/d\rho(r=0)= A $ is the sound speed must be between 0 and 1, so $0 \le A \le 1.  $ 
	Similarly, the expression for $ \frac{dp_{\perp}}{d\rho}> 0 $ at $  r = 0 $ can also be calculated,
	\begin{equation*}
	\frac{dp_{\perp}}{d\rho}_{r=0} =
	\end{equation*}
	
	\begin{equation}
		 \frac{-p_{2}^{-1}(a^{8}nR^{2}(-20A+3n-8An+9A^2n)p_2^{2}  +p_{11}-p_{12} +A^2R^10p_{3}^{2}+a^4p_{13})}{10a^4n(1+n)R^2(a^2+R^2)^2},
	\end{equation}

	\begin{equation*}
		 p_{11}= 2a^{6}np_2R^{4}(2A+A^2(3-6n)-4An+(A^2(9A-3)+n(3-8A)-22A)p_{2}),
	\end{equation*}
	\begin{equation*}
		p_{12} = 2Aa^{2}R^{8}p_{4}\left(A-2An+(2n-A+3An)p_{2}\right),
	\end{equation*}
	
	\begin{equation*}
	 p_{13}= 	A^2R^6\left(1-4n+4n^2+(16n-12n^2-2)p_{2}+(1-12n+9n^2)p_{2}^{2}\right)
	\end{equation*}
	\begin{equation*}
		-4AnR^6p_{2}(-2+2n+(2n+7)p_{2})+3n^2R^6p_{2}^{2}.
	\end{equation*}
	Fig.(\ref{fig.6}) and Fig.(\ref{fig.7}) shows the graphical representation of $ \frac{dp_{r}}{d\rho} $ and $ \frac{dp_{\perp}}{d\rho} $, which lies between 0 and 1.
	\subsection{Energy Condition:}
	\label{sec:7}
	Each of the energy conditions, namely Weak Energy Condition (WEC), Null Energy Condition (NEC), Strong Energy Condition (SEC), Dominant Energy Condition (DEC) and Trace Energy Condition (TEC)  are satisfied for an anisotropic fluid sphere to be physically accepted matter composition if and only if the following inequalities hold simultaneously in every point inside the fluid sphere.
	\begin{equation}
		(i) NEC : \rho + p_{r}\ge 0 ,  \rho + p_{\perp}\ge 0,
	\end{equation}
	\begin{equation}
		(ii) WEC:  \rho + p_{r} > 0 , \rho > 0,
	\end{equation}
	\begin{equation}
		(iii) SEC:  \rho + p_{r} +2 p_{\perp}\ge 0,
	\end{equation}
	\begin{equation}
		(iv) DEC: \rho > \vert p_{r} \vert, \rho > \vert p_{\perp} \vert,
	\end{equation}
	\begin{equation}
		(v) TEC: \rho - p_{r} - 2 p_{\perp}\ge 0.
	\end{equation}
	
	We have 
	\begin{equation*}
		\rho - p_{r} -2 p_{\perp}({r=0}) =
	\end{equation*}
	\begin{equation}
		\frac{-3(a^4n(3A-1)p_{2}+a^2R^2(A-2An+p_{2}(3An-A-n)) -AR^4p_{4})}{a^2R^2(a^2+R^2)p_{2}};
	\end{equation}
	\begin{equation*}
		\rho - p_{r} -2 p_{\perp}({r=a}) =
	\end{equation*}
	
	\begin{equation}
		\frac{p_{14}+p_{2}(a^4(-2-4A)+R^2a^2(2n-8-8A) -4AR^4+p_{2}^{2}(a^4+2a^2+R^4))}{2a^{2}(a^2+R^2)^{2}p_{2}};
	\end{equation}
	where,
	\begin{equation*}
		p_{14}=a^4(3+4A-6n-4An-8An^2)+R^4(Ap_{4})+a^2R^2(6+8A-6n+4An);
	\end{equation*}
	
	As we have shown the graph of density, radial and tangential pressures in Fig. (\ref{fig.3}), Fig.(\ref{fig.4}) and Fig.(\ref{fig.5}). From these figures, we would conclude that density, radial and tangential pressures are monotonically decreasing and positive throught the distribution, so addition of density with pressures are alaways positive. So NEC condition is satisfied throught the distribution. For WEC condition, addition of density with pressure as well as density are positive in the interior of the star. So WEC condition is satisfied. Again summation of density, radial and tangential pressures must be greater than zero in the inside region of the star and it indeed as density, radial and tangential pressures are positive. So SEC condition is satisfied. From the Fig.(\ref{fig.9}) and Fig.(\ref{fig.10}), we have shown the subtraction of pressures from the density is always greater than zero also the value of $ \rho - p_{r}$ and $ \rho - p_{\perp} $ at the boundry of the star is given in Table (\ref{tab:2}). So DEC condition is satisfied. Most important energy condition is TEC, which is satisfied as the value of $ \rho - p_{r} -2 p_{\perp}$ at center as well as boundry of the star is given in the Table (\ref{tab:1}). Fig.(\ref{fig.11}) shows the graphical representation for the star PSR J0348+0432 with different values of 'n'.
	\subsection{Monotony Condition:}
	\label{sec:10}
	A realistic stellar model should have the following
	properties:\\
	\;\;\;\; $\frac{d\rho}{dr}\le 0 ,\;\;\;\;  \frac{dp_{r}}{dr} \le 0 ,\;\;\;\;   \frac{dp_{\perp}}{dr}\le 0  $ \;\;\;\;\;\; for  $ 0 \le r \le a  $
	\\These conditions are checked graphically for the star PSR J0348+0432. Which is shown in Fig.(\ref{fig.12}). Also the values of gradients is provided in Table.(\ref{tab:3}). 
	\subsection{Mass radius relation and surface redshift:}
	Introducing the relation between the mass function $ m(r) $ and the the metric potential $ e^{\lambda}, e^{-\lambda}=1-\frac{2m}{r}$, the
	expression for mass function can be obtained as, 
	\begin{equation}
		m(r)=\int_{0}^{r}4\pi\rho r^2 \,dr = \frac{r}{2}\left(1-\frac{1}{(1-\frac{r^2}{R^2})^n}\right),
	\end{equation}
	the compactness u(r) is obtained from the formula $ u(r)= \frac{m(r)}{r} =\frac{1}{2}\left(1-\frac{1}{(1-\frac{r^2}{R^2})^n}\right). $
	\\The surface redshift $ z_s $can be obtained as,
	\begin{equation}
		z_s= \left(1-2\frac{m(r)}{r}\right)^{-\frac{1}{2}}-1.
	\end{equation}
	The gravitational red-shift of the stellar configuration is given by
	\begin{equation}
		z= e^{-\nu/2}-1.  
	\end{equation}
	The profiles of mass function, compactness and surface redshift are shown in Fig.(\ref{fig.13}), Fig.(\ref{fig.14}), Fig.(\ref{fig.15}) for different values of 'n' mentioned in the figure. All functions are monotonic increasing functions of ‘r’. Surface redshift can be used to explain the strong
	physical interaction between the internal particles of the star and its equation of state. In our context, according to Buchdahl \cite{buchdahl1959general} compact star	satisfy the Buchdahl requirement $ u(r) < \frac{4}{9}.$ Barraco and Hamity \cite{barraco2002maximum} established that in the absence of the cosmological constant, the value of redshift $z_s$ of an isotropic star must be near 2. Following that, B{\"o}hmer and Harko  \cite{bohmer2006bounds} generalised the previous conclusion for the situation of anisotropy in conjunction with the cosmological constant and discovered that $z_s$ must be less than 5. However, in one of his pioneering articles, Ivanov \cite{ivanov2002maximum} demonstrated that adjustments or restrictions led to the greatest permissible value $\le $ 5.211. We estimated the values of gravitational redshift for various values of 'n' in Table (\ref{tab:1}), indicating that our model predicts a stable compact star. In Fig.(\ref{fig.16}), the profile of gravitational red-shift is shown against r, which decreases monotonically in nature.
	\subsection{Equation of state}
	It is also necessary to discover a link between pressure and density, known as the equation of state. The model is created by solving field equations and assuming a linear relationship between radial pressure and matter density; nevertheless, the relationship between transverse pressure and matter density remains uncertain. In Fig.(\ref{fig.17}), we have depicted the nature of tangential pressure variation with regard to density graphically for the star PSR J0348+0432 with different values of 'n'.
	
	\section{Physical Viability and Stability:}
	\label{sec:6}
	\noindent
	Let us now examine the model's physical viability and stability in light of the following.
	\subsection{Stability under three different forces}
	\label{sec:13}
	A star maintains static equilibrium in the presence of three forces: gravitational force $(F_{g})$, hydrostatic force $ (F_{h}) $, and anisotropic force $(F_{a})$. This condition is mathematically expressed as the Tolman-Oppenheimer-Volkoff  (TOV)(Tolman \cite{tolman1939static}, Oppenheimer and Volkoff \cite{oppenheimer1939massive}) equation, which is described by the conservation equation provided by
	\begin{equation}\label{tov1}
		\nabla^{\mu}T_{\mu\nu}=0.
	\end{equation}
	Now using the expression given in equation (\ref{e4}) into (\ref{tov1}), one can obtain the following
	equation:
	\begin{equation}\label{TOV}
		\frac{-\nu'}{2}(\rho+p_{r})-\frac{dp_{r}}{dr}+\frac{2}{r}(p_{\perp}-p_{r})= 0.
	\end{equation}
	This equation (\ref{TOV}) can be written as
	\begin{equation}
		F_{g}+ F_{h}+F_{a}=0,
	\end{equation}
	where 
	\begin{equation*}
		F_{g} = \frac{-\nu'}{2}(\rho+p_{r}), 
	\end{equation*}
	\begin{equation*}
		F_{h} = -\frac{dp_{r}}{dr},
	\end{equation*}
	\begin{equation*}
		F_{a} = \frac{2}{r}(p_{\perp}-p_{r}). 
	\end{equation*}
	The Fig.(\ref{fig.18}) shows that hydrostatics and anisotropic force are positive, but gravitational force is negative to keep the system in static equilibrium.
	\subsection{Adiabatic index for stability}
	The adiabatic index which is defined as
	\begin{equation}\label{adiabatic}
		\Gamma =\left(\frac{\rho +p_{r}}{p_{r}}\right) \frac{dp_{r}}{d\rho}
	\end{equation}
	is related to the stability of a relativistic anisotropic stellar configuration.
	Any stellar configuration will remain stable if the adiabatic index is greater than 4/3, which physically characterises the stiffness of the EoS for a given density. Chandrasekhar \cite{chandrasekhar1964erratum} was the first to investigate this, and he demonstrated, using Eq.(\ref{adiabatic}) in his work, that under general relativity, the Newtonian lower limit $\frac{4}{3} $ has a significant effect. Later, several other researcher  Heintzmann and  Hillebrandt \cite{heintzmann1975neutron}, Hillebrandt and  Steinmetz \cite{hillebrandt1976anisotropic}, Barreto et al.\cite{barreto1992generalization}, Chan et al.\cite{chan1993dynamical}, Doneva and Yazadjiev \cite{doneva2012nonradial}, Moustakidis \cite{moustakidis2017stability} investigated the adiabatic index within a dynamically stable stellar system in the presence of an infinitesimal radial adiabatic perturbation.
	In Fig.(\ref{fig.19}), we have shown the nature of Relativistic adiabatic index variation graphically for the star PSR J0348+0432 with different values of 'n'. Moreover the profile is monotonic increasing as well as greater than 4/3 everywhere inside the stellar interior.
	\subsection{Herrera condition for stability}
	We also know that the velocity of sound (both radial and transverse) must be less than the speed of light for a physically viable model, which is known as the causality requirement. 
	The velocity of sound, which is given by $ v_{s}^2 = \frac{dp}{d\rho} $, is one method for quantifying dense materials. 
	where p is the pressure and  $ \rho $ is the matter density.
	According to Lattimer and Prakash \cite{lattimer2001neutron} causality requires an absolute limit on
	$ v_{s}^2 \le 1 $, and thermodynamic stability ensures that $ v_{s}^2 \ge 0. $ 
	There are two possible stable and two possible unstable regions: one with a positive gradient of anisotropy with respect to the radial variable r and other with a negative gradient. According to Abreu et al.\cite{abreu2007sound}, it is obvious $ 0\le \frac{dp_{r}}{dr} \le 1 ,\;\;   0 \le \frac{dp_{\perp}}{dr}\le 1.  $ 
	i.e $ 0 \le \vert v_{s\perp}^2-v_{sr}^2 \vert\le 1. $

	\begin{equation}
	-1 \le \vert v_{s\perp}^2-v_{sr}^2 \vert\le 1 \Rightarrow	\left\{
		\begin{array}{rcl}
			\ -1 \le \vert v_{s\perp}^2-v_{sr}^2 \vert\le 0 & \mbox{potentially stable} \\
			0 \le \vert v_{s\perp}^2-v_{sr}^2 \vert\le 1 & \mbox{potentially unstable} 
		\end{array} \right.
	\end{equation}
	
	As a result, we may now investigate possible stable/unstable categories inside anisotropic models based on differences in sound propagation within the matter configuration. Those areas $ v_{sr}^2 \ge  v_{s\perp}^2 $ that will be stable. On the other hand, if $ v_{sr}^2 \le  v_{s\perp}^2 $ everywhere within a matter distribution, no cracking will occur. From Fig.(\ref{fig.6}) we can observe that the graph of $ \frac{dp_r}{d\rho} $ is straight line at 0.2, which indicates $\frac{dp_r}{d\rho} = A  $ and A is taken 0.2, which is lies between 0 and 1. From Fig.(\ref{fig.7}) we observed that the graph of  $\frac{dp_{\perp}}{d\rho} $, which is lies in range of 0 to 1. It is also noticed that from the Fig.(\ref{fig.20}), the matter configuration relation is stable as aforementioned since  $ v_{s\perp}^2 - v_{sr}^2 \le 0 $ everywhere inside the star.  
	\section{Discussion:}
	\label{sec:11}
\begin{figure}[H]\centering
		\includegraphics[scale = 1]{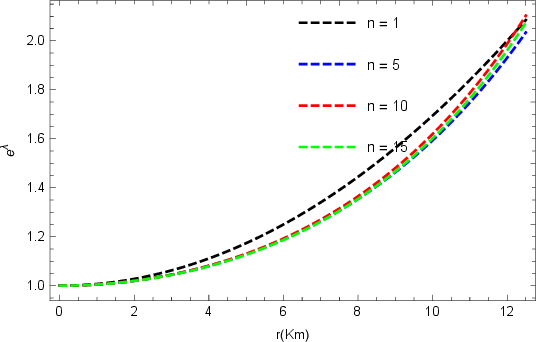} 
		\caption{The metric potentials $e^{\lambda} $ is plotted against r inside the stellar interior.}
		\label{fig.1}
	\end{figure}

\begin{figure}[H]\centering
		\includegraphics[scale = 1]{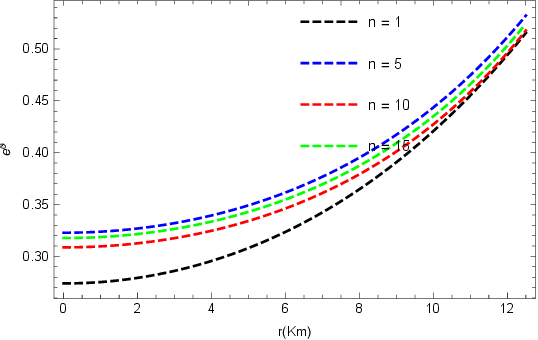} 
		\caption{The metric potentials $e^{\nu} $ is plotted against r inside the stellar interior.}
		\label{fig.2}
	\end{figure}

\begin{figure}[H]\centering
		\includegraphics[scale = 1]{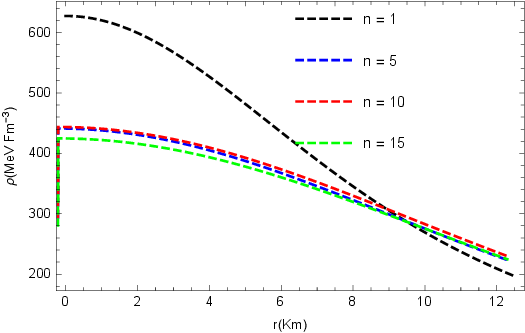} 
		\caption{Variation of density against radial variable $r$.}
		\label{fig.3}
	\end{figure}

\begin{figure}[H]\centering
		\includegraphics[scale = 1]{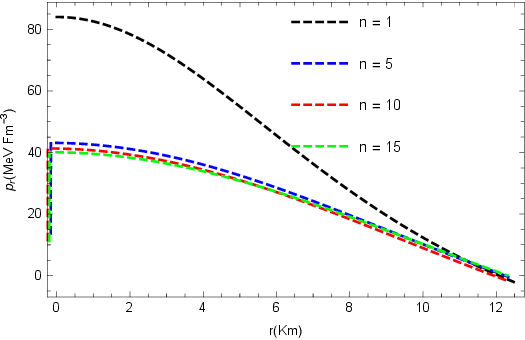}
		\caption{Variation of radial pressures against radial variable $r$.}
		\label{fig.4}
	\end{figure}

\begin{figure}[H]\centering
		\includegraphics[scale = 1]{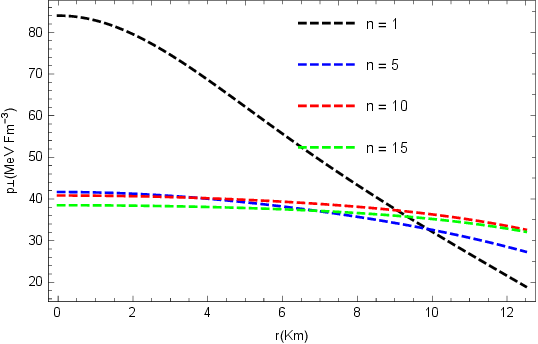}
		\caption{Variation of transverse pressures against radial variable $r$ 
		}
		\label{fig.5}
	\end{figure}

\begin{figure}[H]\centering
		\includegraphics[scale = 1.0]{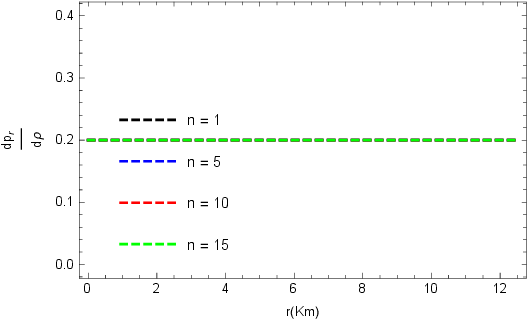}
		\caption{Variation of $ \frac{dp_r}{d\rho} $ against radial variable $r$. 
		}
		\label{fig.6}
	\end{figure}

\begin{figure}[H]\centering
		\includegraphics[scale=1]{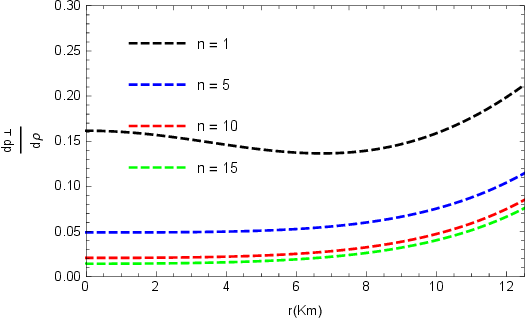}
		\caption{Variation of $ \frac{dp_\perp}{d\rho} $ against radial variable $r$.
		}
		\label{fig.7}
	\end{figure}

\begin{figure}[H]\centering
		\includegraphics[scale = 1]{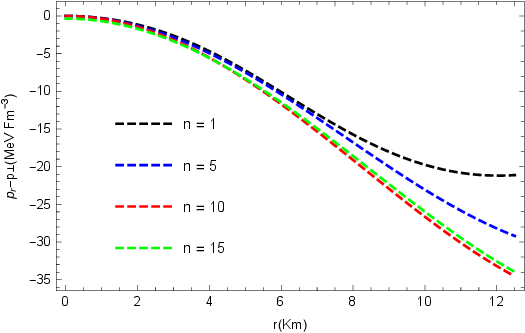}
		\caption{Variation of anisotropies against radial variable $r$. 
		}
		\label{fig.8}
	\end{figure}

\begin{figure}[H]\centering
		\includegraphics[scale = 1]{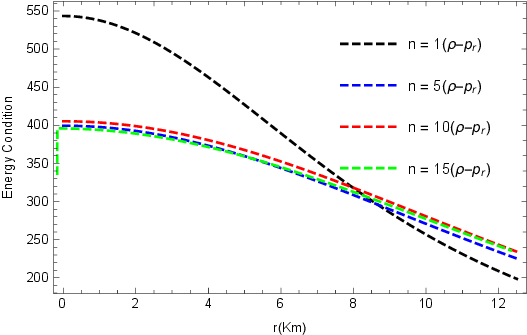} 
		\caption{ $ \rho-p_r $
			is plotted against r inside the stellar interior for the compact star  PSR J0348+0432.}
		\label{fig.9}
	\end{figure}

\begin{figure}[H]\centering
		\includegraphics[scale = 1]{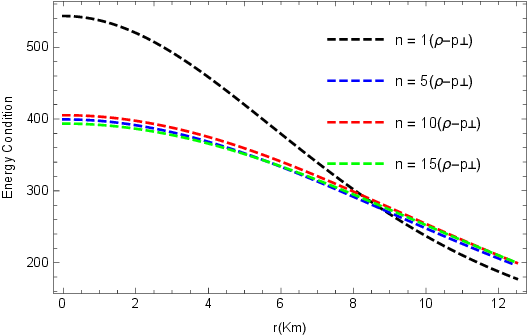} 
		\caption{ $ \rho-p_{\perp} $
			is plotted against r inside the stellar interior for the compact star  PSR J0348+0432.}
		\label{fig.10}
	\end{figure}

\begin{figure}[H]\centering
		\includegraphics[scale = 1]{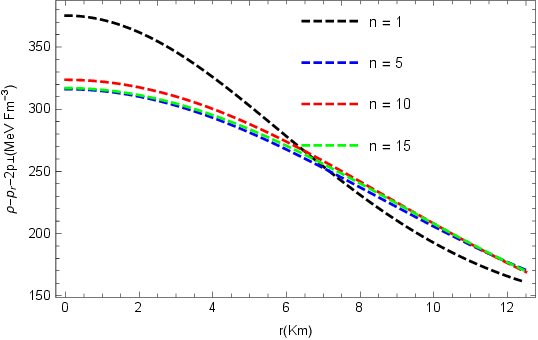}
		\caption{Variation of trace energy condition against radial variable $r$. 
		}
		\label{fig.11}
	\end{figure}

\begin{figure}[H]\centering
		\includegraphics[scale = 1]{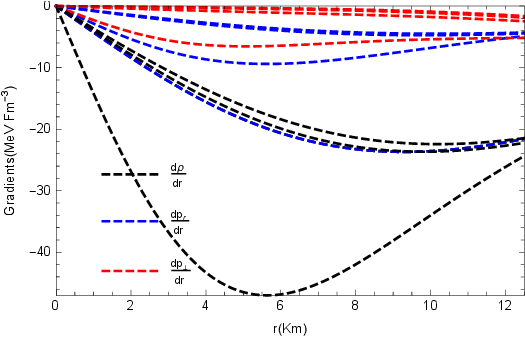}
		\caption{Variation of a Gradients $ \frac{d\rho}{dr} $,$ \frac{dp_{r}}{dr} $ and $ \frac{dp_{\perp}}{dr} $ with respect
			to the radial coordinate r.
		}
		\label{fig.12}
	\end{figure}

\begin{figure}[H]\centering
		\includegraphics[scale = 1]{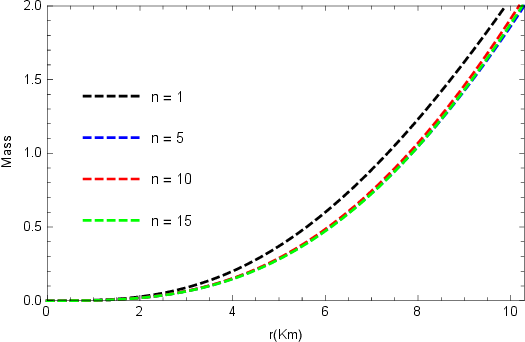}
		\caption{Variation of mass  with the radial coordinate r for compact star PSR J0348+0432. 
		}
		\label{fig.13}
	\end{figure}

\begin{figure}[H]\centering
		\includegraphics[scale = 1]{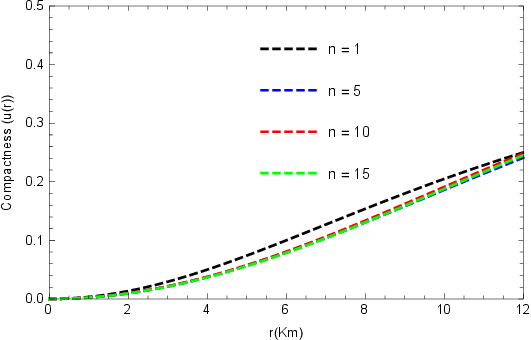}
		\caption{Compactness  with the radial coordinate r for compact star PSR J0348+0432. 
		}
		\label{fig.14}
	\end{figure}

\begin{figure}[H]\centering
		\includegraphics[scale = 1]{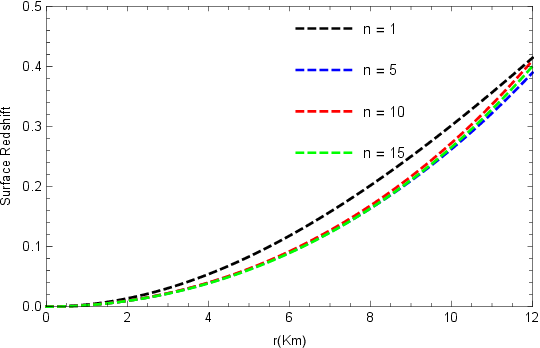}
		\caption{Surface redshift  with the radial coordinate r for compact star PSR J0348+0432. 
		}
		\label{fig.15}
	\end{figure}

\begin{figure}[H]\centering
		\includegraphics[scale = 1.0]{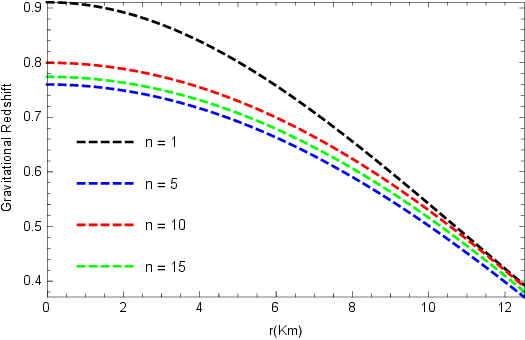}
		\caption{Gravitational redshift function is plotted against r inside the stellar interior for the	compact star PSR J0348+0432. 
		}
		\label{fig.16}
	\end{figure}

\begin{figure}[H]\centering
		\includegraphics[scale = 1]{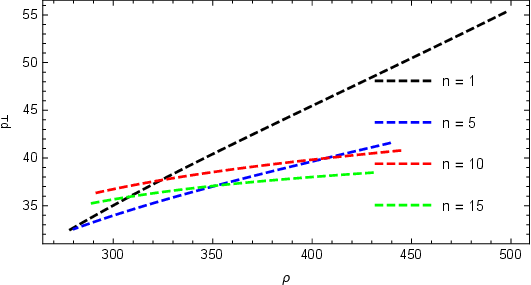}
		\caption{The relation between the pressure $p_{\perp} $ and density $\rho $  is plotted  for the compact star PSR J0348+0432. 
		}
		\label{fig.17}
	\end{figure}

\begin{figure}[H]\centering
		\includegraphics[scale = 1]{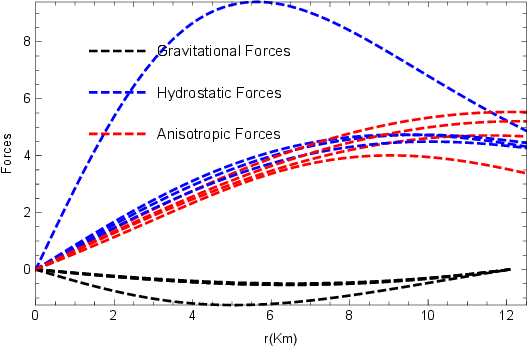}
		\caption{Variation of three forces like Gravitational Force(Black),Hydrostatic Force(Blue) and Anisotropic Force(Red) with respect
			to the radial coordinate r. 
		}
		\label{fig.18}
	\end{figure}

\begin{figure}[H]\centering
		\includegraphics[scale = 1]{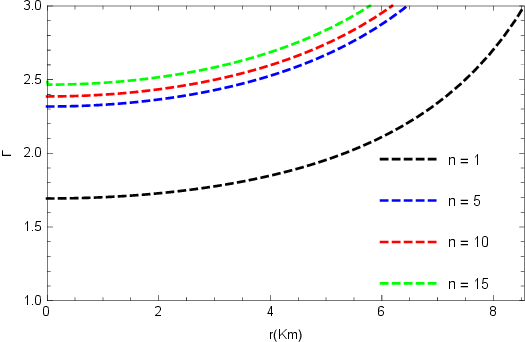}
		\caption{Variation of adiabatic Index against radial variable $r$. 
		}
		\label{fig.19}
	\end{figure}

\begin{figure}[H]\centering
		\includegraphics[scale = 1]{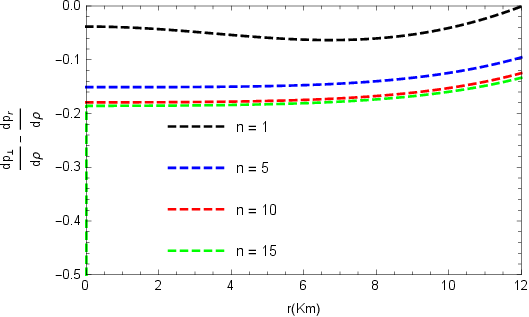} 
		\caption{The causality
			condition with respect to the radial coordinate r.}
		\label{fig.20}
	\end{figure}

	\noindent
	We derived a class of interior solutions to the Einstein field equations for an anisotropic matter distribution obeying a linear EoS in this study. The solution appears to be interesting because it is regular and well behaved, and so potentially explain a relativistic compact star.
	To demonstrate that the solution is applicable to compact observable sources, we consider the pulsar PSR J0348+0432, with the mass  $ 2.01^{+ 0.03}_{- 0.03} M_\odot $ and Radius $ R = 12.072  km $ respectively by Zhao \cite{zhao2015properties}. For the given mass and
	radius, we have chosen the values of the constants $ A = 0.2 $. We attempted to figure out the behaviour of the physically important quantities graphically within the stellar interior using the values of the constants and plugging in the values of A, R and C for the physical acceptability of our model. As a result, based on the graphical plots, which typically illustrate the basic properties of a given model, we would like to highlight the following salient features of our proposed model:
	\\ Fig.(\ref{fig.1}) and Fig.(\ref{fig.2}) demonstrates that the metric potentials are positive within the stellar interior, as required. In Fig.(\ref{fig.3}), Fig.(\ref{fig.4}) and Fig.(\ref{fig.5}) shows variations in energy density, radial pressure and tangential pressure. Pressures decrease radially outwards from their highest value at the centre and radial pressure drops to zero at the boundary, as expected, while tangential pressure stays nonzero at the boundary. All of the quantities, obviously, drop monotonically from the centre to the boundary. Fig.(\ref{fig.6}) and Fig.(\ref{fig.7}) depicts the satisfaction of the causality criterion. Fig.(\ref{fig.8}) illustrates the anisotropy variation as the surface anisotropy decreases. Fig.(\ref{fig.9}) and Fig.(\ref{fig.10}) shows dominent energy condition and its satisfied throughout the interior of the star. Fig.(\ref{fig.11}) shows that trace energy condition $\rho-p_{r}-2p_{\perp}> 0 $ is satisfied throughout the distribution. As $\rho-p_{r}-2p_{\perp}> 0 $ indicates that $ \rho > p_{r} $ and $\rho > p_{\perp} $, which shows that all the Energy conditions NEC, WEC, SEC are satisfied. Variation of the gradient of these physical parameters $\frac{d\rho}{dr}, \frac{dp_{r}}{dr}, \frac{dp_{\perp}}{dr} $ indicates in Fig.(\ref{fig.12}).  Fig.(\ref{fig.13}) reprsentaion of mass against radius r, which is increasing throught the stellar interior. Fig.(\ref{fig.14}) and Fig.(\ref{fig.15}) reprsentaion of compactness and surface redshift against radius r, which is increasing throught the stellar interior. Fig.(\ref{fig.16}) shows the graphical repersentation of gravitational redshift. Fig.(\ref{fig.17}) shows equation of state in terms of tangential pressure and density.  Verification of the forces with respect to the radial coordinate r have been depicted with their expected
	unique features in Fig.(\ref{fig.18}). It can be noted that the outwardly acting combined
	anisotropic and hydrostatic forces balance the inwardly acting strong gravitational force. Fig.(\ref{fig.19}) shows the variation of adiabatic index with radius is satisfied throughout the distribution.  Furthermore, within the stellar interior, the stability factor $ v_{\perp}^2-v_{r}^2 $ is always negative, indicating that the solution is stable under anisotropy perturbation, which shown in Fig.(\ref{fig.20}). Hence the model is suitable to describe PSR J0348+0432.

	\begin{table}[ht]\centering
		\caption{The numerical values of the strong energy condtion at center as well as surface, redshift at surface and adiabatic Index at surface for the compact star PSR J0348+0432.}
		\label{tab:1}       
		\begin{tabular}{llllll} 
			\hline\noalign{\smallskip}
			\textbf{n} & 
			{$ \mathbf{ \rho - p_{r} - 2p_{\perp}}_{(r=0)} $} & {$ \mathbf{\rho-p_{r}-2p_{\perp}}_{(r=a)} $} &    {$ \mathbf{ Z_{(r=0)}} $} &    {$ \mathbf{ Z_{(r=a)}} $} &  {$ \mathbf{\Gamma_{(r=0)}}$} 
			
			\\	&   \textbf{(MeV fm{$\mathbf{^{-3}}$})} & \textbf{(MeV fm{$\mathbf{^{-3}}$})} &    \textbf{(Redshift)} &    \textbf{(Redshift)} & \textbf{(Adiabatic }   \\
			& \textbf{}	&  \textbf{} & \textbf{} & \textbf{} &\textbf{ Index)}   \\
			\noalign{\smallskip}\hline\noalign{\smallskip}
			\textbf{$ n = 1 $} 	  & 375.327  & 164.929 &    0.9105   & 0.4183 &1.69 \\
			\textbf{$ n = 2 $} 	  & 333.673   & 172.778 &   0.8129   & 0.4035& 2.00\\
			\textbf{$ n = 3 $} 	  & 325.55    & 174.541  & 0.793 & 0.406  &2.14\\
			\textbf{$ n = 4 $}    & 324.299   & 174.98 &  0.792  & 0.402 &2.22 \\
			\textbf{$ n = 5 $}    & 316.126   & 176.10 &   0.759   & 0.394 & 2.31\\
			\textbf{$ n = 6 $}    & 318.422   & 175.939  & 0.772   & 0.400 &2.33\\
			\textbf{$ n = 7 $}    & 317.108  & 176.147  &   0.767    & 0.399 &2.37\\
			\textbf{$ n = 8 $} 	  & 325.411  & 174.952 &    0.806   & 0.417 &2.34 \\
			\textbf{$ n = 9 $} 	  & 319.635  & 175.86 &   0.780   & 0.406 & 2.39\\
			\textbf{$ n = 10 $}  & 323.665   & 175.235  & 0.799 & 0.415& 2.38\\
			\textbf{$ n = 11 $}    & 316.031   & 176.354 &  0.764   & 0.401 &2.44\\
			\textbf{$ n = 12 $}    & 317.687   & 176.144 &   0.772   & 0.404 & 2.56 \\
			\textbf{$ n = 13 $}    & 318.398   & 176.05  & 0.776    & 0.406& 2.44\\
			\textbf{$ n = 14 $}    & 318.305   & 176.04  &   0.777    & 0.406&2.82\\
			\textbf{$ n = 15 $}  & 316.995    & 175.864  & 0.774 & 0.405& 2.88 \\

			\noalign{\smallskip}\hline
		\end{tabular} 
		
	\end{table}
	\begin{table}[ht]\centering
		\caption{The numerical values of the $ \frac{dp_{r}}{d\rho} $ at center as well as surface and $ \frac{dp_{\perp}}{d\rho} $ at center as well as surface for the compact star PSR J0348+0432.}
		\label{tab:2}
		\begin{tabular}{llllllll}
			\hline\noalign{\smallskip}
			\textbf{n} &  {$ \mathbf{\frac{dp_{r}}{d\rho}_{(r=0)}} $} & {$ \mathbf{\frac{dp_{\perp}}{d\rho}_{(r=0)}} $}  & {$ \mathbf{\frac{dp_{r}}{d\rho}_{(r=a)}} $} & {$ \mathbf{\frac{dp_{\perp}}{d\rho}_{(r=a)}} $} & {$ \mathbf{(\nu^{2}_{t}-\nu^{2}_{r})_{(r=a)}} $}& {$ \mathbf{(\rho-p_{r})_{(r=a)}} $}&  {$ \mathbf{(\rho-p_{\perp})_{(r=a)}} $}\\
			&   \textbf{} & \textbf{} & \textbf{}   \\
			\noalign{\smallskip}\hline\noalign{\smallskip}
			\\\textbf{$ n = 1 $} 	  & 0.2  & 0.1612 &    0.2   & 0.1913 & -0.0087 & 242.02 & 208.63 \\
			\textbf{$ n = 2 $} 	  & 0.2   & 0.103 &    0.2  & 0.157 & -0.043 & 222.45 & 197.61\\
			\textbf{$ n = 3 $} 	  & 0.2    & 0.075  &  0.2  & 0.1323 & -0.0677 & 228.9 & 201.72\\
			\textbf{$ n = 4 $}    & 0.2   & 0.0581 &   0.2    & 0.115  & -0.085 & 233.21 & 204.09 \\
			\textbf{$ n = 5 $}    & 0.2   & 0.0488 &    0.2    & 0.105  & -0.095 & 232.79 & 204.45\\
			\textbf{$ n = 6 $}    & 0.2   & 0.0399  &  0.2    & 0.096   & -0.104  & 235.71 & 205.82\\
			\textbf{$ n = 7 $}    & 0.2   & 0.0342  &    0.2     & 0.0904  & -0.1096 & 236.55 & 206.35\\
			\textbf{$ n = 8 $} 	  & 0.2  & 0.0269&     0.2    & 0.0835 & -0.1165 &241.21 & 208.08 \\
			\textbf{$ n = 9 $} 	  & 0.2   & 0.0249 &    0.2    & 0.0806 & -0.1194 & 239.57 & 207.76\\
			\textbf{$ n = 10 $}  & 0.2   & 0.0206 &  0.2  & 0.0762 & -0.1238 & 242.02 & 208.63\\
			\textbf{$ n = 11 $}    & 0.2   & 0.0204 &   0.2   & 0.0751 & -0.1249 & 239.17 & 207.71\\
			\textbf{$ n = 12 $}    & 0.2   & 0.0021 &    0.2    & 0.0722 & -0.1278 & 240.38 & 208.26 \\
			\textbf{$ n = 13 $}    & 0.2   & 0.0006  &  0.2     & 0.0699 & -0.1301 & 241.10 & 208.57\\
			\textbf{$ n = 14 $}    & 0.2   & 0.0001  &    0.2     & 0.0681 & -0.1319 & 241.41 & 208.72\\
			\textbf{$ n = 15 $}  & 0.2    & 0.0004  &  0.2  & 0.0675 & -0.1325 & 241.41 & 208.63 \\

			\noalign{\smallskip}\hline
		\end{tabular} 
		
	\end{table}
	\newpage
	\begin{table}[H]\centering
		\caption{The numerical values of the $ \frac{d\rho}{dr} $,$ \frac{dp_{r}}{dr} $ and  $ \frac{dp_{\perp}}{dr} $ at center as well as surface for the compact star PSR J0348+0432.}
		\label{tab:3}
		\begin{tabular}{lllllll}
			\hline\noalign{\smallskip}
			\textbf{n} &  {$ \mathbf{\frac{d\rho}{dr}_{(r=0)}} $} & {$ \mathbf{\frac{d\rho}{dr}_{(r=a)}} $}  & {$ \mathbf{\frac{dp_{r}}{dr}_{(r=0)}} $}& {$ \mathbf{\frac{dp_{r}}{dr}_{(r=a)}} $} & {$ \mathbf{\frac{dp_{\perp}}{dr}_{(r=0)}} $} & {$ \mathbf{\frac{dp_{\perp}}{dr}_{(r=a)}} $}\\
			&   \textbf{} & \textbf{} & \textbf{}   \\
			\noalign{\smallskip}\hline\noalign{\smallskip}
			\\\textbf{$ n = 1 $} 	  & 0  & -25.887   &    0   & -5.177 & 0& -5.195\\
			\textbf{$ n = 2 $} 	  & 0   & -24.11     &    0  & -4.822 & 0&-3.790\\
			\textbf{$ n = 3 $} 	  & 0    & -23.32    &  0  & -4.66 & 0& -3.08\\
			\textbf{$ n = 4 $}    & 0   & -23.085 &   0    & -4.617  & 0& -2.677 \\
			\textbf{$ n = 5 $}    & 0   & -22.063 &    0    & -4.412  & 0& -2.327\\
			\textbf{$ n = 6 $}    & 0   & -22.229  &  0    & -4.445   & 0&-2.150\\
			\textbf{$ n = 7 $}    & 0   & -22.001  &    0     & -4.400  & 0&-1.990\\
			\textbf{$ n = 8 $} 	  & 0  & -22.912 &     0    & -4.582 & 0& -1.915\\
			\textbf{$ n = 9 $} 	  & 0   & -22.184 &    0    & -4.436 & 0 & -1.788\\
			\textbf{$ n = 10 $}  & 0   & -22.620 &  0  & -4.524 & 0 & -1.724\\
			\textbf{$ n = 11 $}    & 0   & -21.679 &   0   & -4.335 & 0 & -1.628\\
			\textbf{$ n = 12 $}    & 0   & -21.84 &    0    & -4.368 & 0 & -1.578\\
			\textbf{$ n = 13 $}    & 0   & -21.901  &  0     & -4.380 & 0 &-1.532\\
			\textbf{$ n = 14 $}    & 0   & -21.87  &    0     & -4.374 & 0 &-1.491\\
			\textbf{$ n = 15 $}  & 0    & -21.77  &  0  & -4.354 & 0 &-1.470 \\

			\noalign{\smallskip}\hline
		\end{tabular} 
		
	\end{table}
	
	
	
	
	

	
	\pagebreak
	\newpage
	\section{References}
	
	\bibliographystyle{unsrt}
	\bibliography{Manuscript.bib}

\begin{thebibliography}{10}

\bibitem{ruderman1972pulsars}
M~Ruderman.
\newblock Pulsars: structure and dynamics.
\newblock {\em Annual Review of Astronomy and Astrophysics}, 10(1):427--476,
  1972.

\bibitem{bowers1974anisotropic}
Richard~L Bowers and EPT Liang.
\newblock Anisotropic spheres in general relativity.
\newblock {\em The Astrophysical Journal}, 188:657, 1974.

\bibitem{sokolov1980phase}
AI~Sokolov.
\newblock Phase transitions in a superfluid neutron liquid.
\newblock {\em Sov. Phys. JETP}, 52(4):575, 1980.

\bibitem{kippenhahn1990stellar}
Rudolf Kippenhahn, Alfred Weigert, and Achim Weiss.
\newblock {\em Stellar structure and evolution}, volume 192.
\newblock Springer, 1990.

\bibitem{sawyer1972condensed}
Raymond~F Sawyer.
\newblock Condensed $\pi$- phase in neutron-star matter.
\newblock {\em Physical Review Letters}, 29(6):382, 1972.

\bibitem{herrera1997local}
Luis Herrera and Nilton~O Santos.
\newblock Local anisotropy in self-gravitating systems.
\newblock {\em Physics Reports}, 286(2):53--130, 1997.

\bibitem{herrera39santos}
L~Herrera.
\newblock Santos. no (1998).
\newblock {\em J. Math. Phys}, 39:3817.

\bibitem{herrera1985isotropic}
L~Herrera and J~Ponce~de Leon.
\newblock Isotropic and anisotropic charged spheres admitting a one-parameter
  group of conformal motions.
\newblock {\em Journal of mathematical physics}, 26(9):2302--2307, 1985.

\bibitem{dev2002anisotropic}
Krsna Dev and Marcelo Gleiser.
\newblock Anisotropic stars: exact solutions.
\newblock {\em General relativity and gravitation}, 34:1793--1818, 2002.

\bibitem{dev2003anisotropic}
Krsna Dev and Marcelo Gleiser.
\newblock Anisotropic stars ii: stability.
\newblock {\em arXiv preprint gr-qc/0303077}, 2003.

\bibitem{cosenza1982evolution}
M~Cosenza, L~Herrera, M~Esculpi, and L~Witten.
\newblock Evolution of radiating anisotropic spheres in general relativity.
\newblock {\em Physical Review D}, 25(10):2527, 1982.

\bibitem{bayin1982anisotropic}
Sel{\c{c}}uk~{\c{S}} Bayin.
\newblock Anisotropic fluid spheres in general relativity.
\newblock {\em Physical Review D}, 26(6):1262, 1982.

\bibitem{krori1984some}
KD~Krori, P~Borgohain, and Ranjumani Devi.
\newblock Some exact anisotropic solutions in general relativity.
\newblock {\em Canadian journal of physics}, 62(3):239--246, 1984.

\bibitem{ponce1987new}
J~Ponce~de Leon.
\newblock New analytical models for anisotropic spheres in general relativity.
\newblock {\em Journal of mathematical physics}, 28(5):1114--1117, 1987.

\bibitem{bondi1992anisotropic}
Hermann Bondi.
\newblock Anisotropic spheres in general relativity.
\newblock {\em Monthly Notices of the Royal Astronomical Society},
  259(2):365--368, 1992.

\bibitem{herrera2001conformally}
L~Herrera, A~Di Prisco, J~Ospino, and E~Fuenmayor.
\newblock Conformally flat anisotropic spheres in general relativity.
\newblock {\em Journal of Mathematical Physics}, 42(5):2129--2143, 2001.

\bibitem{herrera2009expansion}
L~Herrera, G{\'e}rard Le~Denmat, and Nilton~O Santos.
\newblock Expansion-free evolving spheres must have inhomogeneous energy
  density distributions.
\newblock {\em Physical Review D}, 79(8):087505, 2009.

\bibitem{herrera2008all}
L~Herrera, J~Ospino, and A~Di~Prisco.
\newblock All static spherically symmetric anisotropic solutions of
  einstein’s equations.
\newblock {\em Physical Review D}, 77(2):027502, 2008.

\bibitem{herrera2010cavity}
L~Herrera, G{\'e}rard Le~Denmat, and Nilton~O Santos.
\newblock Cavity evolution in relativistic self-gravitating fluids.
\newblock {\em Classical and Quantum Gravity}, 27(13):135017, 2010.

\bibitem{herrera2012dynamical}
L~Herrera, G~Le Denmat, and NO~Santos.
\newblock Dynamical instability and the expansion-free condition.
\newblock {\em General Relativity and Gravitation}, 44:1143--1162, 2012.

\bibitem{mak2002exact}
MK~Mak and T~Harko.
\newblock An exact anisotropic quark star model.
\newblock {\em Chinese journal of astronomy and astrophysics}, 2(3):248, 2002.

\bibitem{herrera2013general}
L~Herrera and W~Barreto.
\newblock General relativistic polytropes for anisotropic matter: The general
  formalism and applications.
\newblock {\em Physical Review D}, 88(8):084022, 2013.

\bibitem{herrera2014conformally}
L~Herrera, A~Di~Prisco, W~Barreto, and J~Ospino.
\newblock Conformally flat polytropes for anisotropic matter.
\newblock {\em General Relativity and Gravitation}, 46:1--16, 2014.

\bibitem{mak2003anisotropic}
MK~Mak and T~Harko.
\newblock Anisotropic stars in general relativity.
\newblock {\em Proceedings of the Royal Society of London. Series A:
  Mathematical, Physical and Engineering Sciences}, 459(2030):393--408, 2003.

\bibitem{viaggiu2009modeling}
Stefano Viaggiu.
\newblock Modeling usual and unusual anisotropic spheres.
\newblock {\em International Journal of Modern Physics D}, 18(02):275--288,
  2009.

\bibitem{sharma2007class}
R~Sharma and SD~Maharaj.
\newblock A class of relativistic stars with a linear equation of state.
\newblock {\em Monthly Notices of the Royal Astronomical Society},
  375(4):1265--1268, 2007.

\bibitem{thirukkanesh2008charged}
S~Thirukkanesh and SD~Maharaj.
\newblock Charged anisotropic matter with a linear equation of state.
\newblock {\em Classical and Quantum Gravity}, 25(23):235001, 2008.

\bibitem{mafa2014charged}
P~Mafa~Takisa, S~Ray, and SD~Maharaj.
\newblock Charged compact objects in the linear regime.
\newblock {\em Astrophysics and Space Science}, 350:733--740, 2014.

\bibitem{harko2016exact}
T~Harko and MK~Mak.
\newblock Exact power series solutions of the structure equations of the
  general relativistic isotropic fluid stars with linear barotropic and
  polytropic equations of state.
\newblock {\em Astrophysics and Space Science}, 361(9):283, 2016.

\bibitem{hansraj2018equilibrium}
Sudan Hansraj and Ayan Banerjee.
\newblock Equilibrium stellar configurations in rastall theory and linear
  equation of state.
\newblock {\em arXiv preprint arXiv:1807.00812}, 2018.

\bibitem{ivanov2020linear}
BV~Ivanov.
\newblock Linear and riccati equations in generating functions for stellar
  models in general relativity.
\newblock {\em The European Physical Journal Plus}, 135(4):1--14, 2020.

\bibitem{rej2021charged}
Pramit Rej and Piyali Bhar.
\newblock Charged strange star in f (r, t) gravity with linear equation of
  state.
\newblock {\em Astrophysics and Space Science}, 366(4):35, 2021.

\bibitem{das2022anisotropic}
Shyam Das, Bikram~Keshari Parida, Koushik Chakraborty, and Saibal Ray.
\newblock Anisotropic compact star with a linear pressure--density
  relationship.
\newblock {\em International Journal of Modern Physics D}, 31(07):2250053,
  2022.

\bibitem{prasad2022anisotropic}
Amit~Kumar Prasad and Jitendra Kumar.
\newblock Anisotropic relativistic fluid spheres with a linear equation of
  state.
\newblock {\em New Astronomy}, 95:101815, 2022.

\bibitem{patel2023new}
Rinkal Patel, BS~Ratanpal, and DM~Pandya.
\newblock New charged anisotropic solution on paraboloidal spacetime.
\newblock {\em arXiv preprint arXiv:2301.10795}, 2023.

\bibitem{ngubelanga2015compact}
Sifiso~A Ngubelanga, Sunil~D Maharaj, and Subharthi Ray.
\newblock Compact stars with quadratic equation of state.
\newblock {\em Astrophysics and Space Science}, 357:1--9, 2015.

\bibitem{varela2010charged}
Victor Varela, Farook Rahaman, Saibal Ray, Koushik Chakraborty, and Mehedi
  Kalam.
\newblock Charged anisotropic matter with linear or nonlinear equation of
  state.
\newblock {\em Physical Review D}, 82(4):044052, 2010.

\bibitem{feroze2011charged}
Tooba Feroze and Azad~A Siddiqui.
\newblock Charged anisotropic matter with quadratic equation of state.
\newblock {\em General Relativity and Gravitation}, 43:1025--1035, 2011.

\bibitem{maharaj2012regular}
SD~Maharaj and P~Mafa~Takisa.
\newblock Regular models with quadratic equation of state.
\newblock {\em General Relativity and Gravitation}, 44:1419--1432, 2012.

\bibitem{sharma2013relativistic}
Ranjan Sharma and BS~Ratanpal.
\newblock Relativistic stellar model admitting a quadratic equation of state.
\newblock {\em International Journal of Modern Physics D}, 22(13):1350074,
  2013.

\bibitem{malaver2014strange}
Manuel Malaver.
\newblock Strange quark star model with quadratic equation of state.
\newblock {\em arXiv preprint arXiv:1407.0760}, 2014.

\bibitem{malaver2020relativistic}
Manuel Malaver and Hamed Daei~Kasmaei.
\newblock Relativistic stellar models with quadratic equation of state.
\newblock {\em International Journal of Mathematical Modelling \&
  Computations}, 10(2 (SPRING)):111--124, 2020.

\bibitem{herrera2013newtonian}
L~Herrera and W~Barreto.
\newblock Newtonian polytropes for anisotropic matter: General framework and
  applications.
\newblock {\em Physical Review D}, 87(8):087303, 2013.

\bibitem{thirukkanesh2012exact}
S~Thirukkanesh and FC~Ragel.
\newblock Exact anisotropic sphere with polytropic equation of state.
\newblock {\em Pramana}, 78:687--696, 2012.

\bibitem{thirukkanesh2013class}
S~Thirukkanesh and FC~Ragel.
\newblock A class of exact strange quark star model.
\newblock {\em Pramana}, 81:275--286, 2013.

\bibitem{takisa2013some}
P~Mafa Takisa and SD~Maharaj.
\newblock Some charged polytropic models.
\newblock {\em General Relativity and Gravitation}, 45:1951--1969, 2013.

\bibitem{singh2022anisotropic}
Ksh~Newton Singh, SK~Maurya, Piyali Bhar, and Riju Nag.
\newblock Anisotropic solution for polytropic stars in 4 d
  einstein--gauss--bonnet gravity.
\newblock {\em The European Physical Journal C}, 82(9):822, 2022.

\bibitem{finch1989realistic}
Michael~R Finch and James~EF Skea.
\newblock A realistic stellar model based on an ansatz of duorah and ray.
\newblock {\em Classical and Quantum Gravity}, 6(4):467, 1989.

\bibitem{kuchowicz1972differential}
B~Kuchowicz.
\newblock Differential conditions for physically meaningful fluid spheres in
  general relativity.
\newblock {\em Physics Letters A}, 38(5):369--370, 1972.

\bibitem{buchdahl1979regular}
HA~Buchdahl.
\newblock Regular general relativistic charged fluid spheres.
\newblock {\em Acta Physica Polonica. Series B}, 10(8):673--685, 1979.

\bibitem{knutsen1988some}
Henning Knutsen.
\newblock Some physical properties and stability of an exact model of a
  relativistic star.
\newblock {\em Astrophysics and space science}, 140(2):385--401, 1988.

\bibitem{murad2015some}
Mohammad~Hassan Murad and Saba Fatema.
\newblock Some new wyman--leibovitz--adler type static relativistic charged
  anisotropic fluid spheres compatible to self-bound stellar modeling.
\newblock {\em The European Physical Journal C}, 75(11):1--21, 2015.

\bibitem{buchdahl1959general}
Hans~A Buchdahl.
\newblock General relativistic fluid spheres.
\newblock {\em Physical Review}, 116(4):1027, 1959.

\bibitem{barraco2002maximum}
Daniel Barraco and Victor~H Hamity.
\newblock Maximum mass of a spherically symmetric isotropic star.
\newblock {\em Physical Review D}, 65(12):124028, 2002.

\bibitem{bohmer2006bounds}
CG~B{\"o}hmer and T~Harko.
\newblock Bounds on the basic physical parameters for anisotropic compact
  general relativistic objects.
\newblock {\em Classical and Quantum Gravity}, 23(22):6479, 2006.

\bibitem{ivanov2002maximum}
Boiko~V Ivanov.
\newblock Maximum bounds on the surface redshift of anisotropic stars.
\newblock {\em Physical Review D}, 65(10):104011, 2002.

\bibitem{tolman1939static}
Richard~C Tolman.
\newblock Static solutions of einstein's field equations for spheres of fluid.
\newblock {\em Physical Review}, 55(4):364, 1939.

\bibitem{oppenheimer1939massive}
J~Robert Oppenheimer and George~M Volkoff.
\newblock On massive neutron cores.
\newblock {\em Physical Review}, 55(4):374, 1939.

\bibitem{chandrasekhar1964erratum}
S~Chandrasekhar.
\newblock Erratum: the dynamical instability of gaseous masses approaching the
  schwarzschild limit in general relativity.
\newblock {\em The Astrophysical Journal}, 140:1342, 1964.

\bibitem{heintzmann1975neutron}
Hillebrandt Heintzmann and W~Hillebrandt.
\newblock Neutron stars with an anisotropic equation of state-mass, redshift,
  and stability.
\newblock {\em Astronomy and Astrophysics}, 38:51--55, 1975.

\bibitem{hillebrandt1976anisotropic}
W~Hillebrandt and KO~Steinmetz.
\newblock Anisotropic neutron star models-stability against radial and
  nonradial pulsations.
\newblock {\em Astronomy and Astrophysics}, 53:283--287, 1976.

\bibitem{barreto1992generalization}
W~Barreto, L~Herrera, and N~Santos.
\newblock A generalization of the concept of adiabatic index for non-adiabatic
  systems.
\newblock {\em Astrophysics and space science}, 187:271--290, 1992.

\bibitem{chan1993dynamical}
R~Chan, L~Herrera, and NO~Santos.
\newblock Dynamical instability for radiating anisotropic collapse.
\newblock {\em Monthly Notices of the Royal Astronomical Society},
  265(3):533--544, 1993.

\bibitem{doneva2012nonradial}
Daniela~D Doneva and Stoytcho~S Yazadjiev.
\newblock Nonradial oscillations of anisotropic neutron stars in the cowling
  approximation.
\newblock {\em Physical Review D}, 85(12):124023, 2012.

\bibitem{moustakidis2017stability}
Ch~C Moustakidis.
\newblock The stability of relativistic stars and the role of the adiabatic
  index.
\newblock {\em General Relativity and Gravitation}, 49:1--21, 2017.

\bibitem{lattimer2001neutron}
JM~Lattimer and M~Prakash.
\newblock Neutron star structure and the equation of state.
\newblock {\em The Astrophysical Journal}, 550(1):426, 2001.

\bibitem{abreu2007sound}
H~Abreu, Hector Hern{\'a}ndez, and Luis~A N{\'u}nez.
\newblock Sound speeds, cracking and the stability of self-gravitating
  anisotropic compact objects.
\newblock {\em Classical and Quantum Gravity}, 24(18):4631, 2007.

\bibitem{zhao2015properties}
Xian-Feng Zhao.
\newblock The properties of the massive neutron star psr j0348+ 0432.
\newblock {\em International Journal of Modern Physics D}, 24(08):1550058,
  2015.

\end{thebibliography}

\end{document}